\pdfoutput = 1
\documentclass[journal,comsoc]{IEEEtran}

\usepackage[T1]{fontenc}
\usepackage{cite}
\usepackage{amsmath}
\usepackage{amssymb}
\usepackage{mathtools}
\usepackage{siunitx}
\DeclareSIUnit\dBm{dBm}
\usepackage{bm}
\usepackage{commath}
\usepackage{enumitem}
\usepackage{adjustbox}
\usepackage{array}
\usepackage{nth}

\usepackage[cmintegrals]{newtxmath}
\usepackage[caption=false,font=footnotesize]{subfig}
\usepackage{url}

\usepackage{blindtext}

\newcommand{\vect}[1]{\bm{#1}}
\newcommand{\trans}{^{\mathrm{T}}}
\newcommand{\herm}{^{\mathrm{H}}}

\DeclareMathOperator{\e}{e}
\DeclareMathOperator{\jj}{j}

\DeclareMathOperator{\atan2}{atan2}

\let\originalleft\left
\let\originalright\right
\renewcommand{\left}{\mathopen{}\mathclose\bgroup\originalleft}
\renewcommand{\right}{\aftergroup\egroup\originalright}

\usepackage{tikz}
\usepackage{pgfplots}
\usepackage{pgfplotstable}
\pgfplotsset{compat=newest}
\usepgfplotslibrary{colormaps, groupplots, patchplots}
\usetikzlibrary{arrows.meta, backgrounds, calc, decorations, decorations.markings, decorations.pathreplacing, math, patterns, pgfplots.polar, shapes}
\pgfdeclarelayer{bg}    
\pgfsetlayers{bg,main}  

\usepackage{tkz-euclide}

\usepackage[acronym, automake]{glossaries-prefix}
\setacronymstyle{long-short}
\newacronym{1D}{1D}{one-dimensional}
\newacronym{2D}{2D}{two-dimensional}
\newacronym{3D}{3D}{three-dimensional}
\newacronym{3GPP}{3GPP}{Third Generation Partnership Project}
\newacronym{5G}{5G}{fifth generation}
\newacronym{ADCGM}{ADCGM}{Alternating Descent Conditional Gradient Method}
\newacronym[longplural=angles of arrival]{AOA}{AOA}{angle of arrival}
\newacronym[longplural=angles of departure]{AOD}{AOD}{angle of departure}
\newacronym{AWGN}{AWGN}{additive white Gaussian noise}
\newacronym{BS}{BS}{base station}
\newacronym{cdf}{cdf}{cumulative distribution function}
\newacronym{CP}{CP}{cyclic prefix}
\newacronym{CRLB}{CRLB}{Cram\'er-Rao lower bound}
\newacronym{DL}{DL}{downlink}
\newacronym{EFIM}{EFIM}{equivalent FIM}
\newacronym{ESPEB}{ESPEB}{expected \glsentrytext{SPEB}}
\newacronym{EV}{EV}{eigenvalue}
\newacronym{EXIP}{EXIP}{extended invariance principle}
\newacronym[longplural=Fisher information matrices]{FIM}{FIM}{Fisher information matrix}
\newacronym{IoT}{IoT}{internet of things}
\newacronym{LOS}{LOS}{line-of-sight}
\newacronym{MIMO}{MIMO}{multiple-input multiple output}
\newacronym{mm-Wave}{mm-Wave}{millimeter-wave}
\newacronym{NLOS}{NLOS}{non-LOS}
\newacronym{OFDM}{OFDM}{orthogonal frequency-division multiplexing}
\newacronym{OTDOA}{OTDOA}{observed time difference of arrival}
\newacronym{pdf}{pdf}{probability density function}
\newacronym{PEB}{PEB}{position error bound}
\newacronym{pmf}{pmf}{probability mass function}
\newacronym{PSD}{PSD}{positive semidefiniteness}
\newacronym{RCRLB}{RCRLB}{root \glsentrytext{CRLB}}
\newacronym{RE}{RE}{resource element}
\newacronym{RMSE}{RMSE}{root mean square error}
\newacronym{RTT}{RTT}{round-trip time}
\newacronym{Rx}{Rx}{receiver}
\newacronym[prefixfirst={a\ }, prefix={an\ }]{SDP}{SDP}{semidefinite program}
\newacronym{SPEB}{SPEB}{squared position error bound}
\newacronym{SNR}{SNR}{signal-to-noise ratio}
\newacronym[longplural=times of arrival]{TOA}{TOA}{time of arrival}
\newacronym{Tx}{Tx}{transmitter}
\newacronym{UE}{UE}{user equipment}
\newacronym{ULA}{ULA}{uniform linear array}
\newacronym{UCA}{UCA}{uniform circular array}
\newacronym{UL}{UL}{uplink}
\newacronym{UTDOA}{UTDOA}{uplink TDOA}
\newacronym{VA}{VA}{virtual anchor}
\makeglossaries

\newsavebox\glsscratchboxa
\newsavebox\glsscratchboxb
\newsavebox\glsscratchboxc
\newsavebox\glsscratchboxd

\usepackage{algorithm}
\usepackage{algpseudocode}
\algnewcommand{\Input}[1]{%
	\State \textbf{input:} \parbox[t]{.8\linewidth}{\raggedright #1}
}
\algnewcommand{\Output}[1]{%
	\State \textbf{output:} \parbox[t]{.8\linewidth}{\raggedright #1}
}
\algnewcommand{\Initialize}[1]{%
	\State \textbf{initialize:} \parbox[t]{.8\linewidth}{\raggedright #1}
}
\algnewcommand{\Break}{%
	\State \textbf{break}
}

\usepackage[pdfencoding=auto, psdextra]{hyperref}
\usepackage{bookmark}

\title{Position Information from Reflecting Surfaces}

\author{Anastasios~Kakkavas,~\IEEEmembership{Student~Member,~IEEE,}
	Mario~H.~Casta\~neda~Garc\'ia,~\IEEEmembership{Member,~IEEE,}
	Gonzalo~Seco-Granados,~\IEEEmembership{Senior~Member,~IEEE,}
	Henk~Wymeersch,~\IEEEmembership{Senior~Member,~IEEE,}
	Richard~A.~Stirling-Gallacher,~\IEEEmembership{Member,~IEEE,}
	and~Josef~A.~Nossek,~\IEEEmembership{Life~Fellow,~IEEE}
	
	\thanks{This work was supported in part by the EU-H2020 project Fifth Generation Communication Automotive Research and Innovation (5GCAR),
		and in part by the ICREA Academia program and the
		Spanish Ministry of
		Science, Innovation and Universities project TEC2017-89925-R.}
	\thanks{A.~Kakkavas is with the Munich Research Center, Huawei Technologies Duesseldorf GmbH, 80992 Munich, Germany, and also with the Department of Electrical and Computer Engineering, Technical University of Munich, 80333 Munich, Germany (e-mail: anastasios.kakkavas@huawei.com).}
	\thanks{M.~H.~Casta\~{n}eda~Garc\'ia and R.~A.~Stirling-Gallacher are with the Munich Research Center, Huawei Technologies Duesseldorf GmbH, 80992 Munich, Germany (e-mail: mario.castaneda@huawei.com; richard.sg@huawei.com).} 
	\thanks{G. Seco-Granados is with the Department of Telecommunications and Systems Engineering, Universitat Autonoma de Barcelona, Spain (UAB) (e-mail: gonzalo.seco@uab.cat).}
	\thanks{H. Wymeersch is with the Department of Electrical Engineering, Chalmers University of Technology, 412 58 Gothenburg, Sweden (email: henkw@chalmers.se).}
	\thanks{J.~A.~Nossek is with the Department of Electrical and Computer Engineering, Technical University of Munich, 80333 Munich, Germany (e-mail: josef.a.nossek@tum.de).}
}

\hypersetup{
	pdfinfo={
		Title={Position Information from Single-Bounce Reflections},
		Author={Anastasios Kakkavas, Mario H. Casta\~neda Garc\'ia, Gonzalo Seco-Granados, Henk Wymeersch, Richard A. Stirling-Gallacher, Josef A. Nossek}
	}
}

\begin{document}

		
	
	\maketitle
	\tikz[overlay,remember picture]
	{
		
		\node at ($(current page.south west)+(0.62in,0.3cm)$) [rotate=0, anchor=south west] {\parbox{\textwidth}{\footnotesize \footnotesize This work has been submitted to the IEEE for possible publication. Copyright may be transferred without notice, after which this version may no longer be accessible.}};
	}
	
	\begin{abstract}
		In the context of positioning a target with a single-anchor, this contribution focuses on the  Fisher information about the position, orientation and
		clock offset of the target provided by single-bounce reflections. The availability of prior knowledge of the target's environment is taken into account via a prior distribution of the position of virtual anchors, and the rank, intensity and direction of provided information is studied. We show that when no prior knowledge is available, single-bounce reflections offer position information in the direction parallel to the reflecting surface, irrespective of the target's and anchor's locations. We provide a geometrically intuitive explanation of the results and present numerical examples demonstrating their potential implications.
	\end{abstract}


	\IEEEpeerreviewmaketitle
	
	\section{Introduction}
		\label{sec:introduction}
		
		Although the majority of practical positioning systems rely heavily or even exclusively on \gls{LOS} propagation, the role of \gls{NLOS} propagation in wireless positioning has been widely studied. Traditionally, the focus has been on the mitigation of the negative impact of \gls{NLOS} paths on positioning accuracy~\cite{CZ05}, with some approaches completely disregarding \gls{NLOS} links and others aiming to correct the \gls{NLOS}-induced bias in the range estimate~\cite{MGW+10}. An alternative approach is to treat the \gls{NLOS} paths as additional sources of position information. An early work in this direction was~\cite{MYJ07}, where it was shown that given distance, \gls{AOD} and \gls{AOA} measurements of a single-bounce \gls{NLOS} path, the \gls{Rx} can lie on a line segment, and an algorithm exploiting this observation was presented. A similar approach for mobile targets was presented in~\cite{PS08}.
		
		Such approaches became much more relevant for \gls{5G} networks~\cite{WML+16}. The upcoming exploitation of the large chunks of available bandwidth at \gls{mm-Wave} frequencies and the use of antenna arrays with a large number of elements, 
		enable the possibility of highly accurate temporal and angular measurements, and improve the separability of multipath components~\cite{AZA+18}. The increased temporal and angular resolution has made single-anchor positioning~\cite{SGD+15} an attractive option when links to multiple anchors may not be available. Algorithms for single-anchor localization and mapping with a single snapshot have been presented in~\cite{SGD+18,MWB18} and~\cite{TKL+19} among others.
		In~\cite{MWB+19} it was shown that, in a \gls{2D} setup, the set of \gls{TOA}, \gls{AOD} and \gls{AOA} measurements from a single-bounce reflection offers rank-1 information for a receiver with unknown position and orientation. The corresponding eigenvalue of the position and orientation \gls{EFIM} was computed analytically, showing that all 3 measurements are required for extracting additional position information from \gls{NLOS} components. 
		In~\cite{KCS+19} it was shown that single-bounce \gls{NLOS} components can be helpful in resolving the clock offset between an imperfectly synchronized \gls{Tx}-\gls{Rx} pair, allowing for accurate single-anchor positioning.
		
		In this letter, considering flat reflecting surfaces, which we refer to as reflectors, we extend the work of~\cite{MYJ07}, as well as~\cite{MWB+19} and~\cite{KCS+19}, as follows:
		\begin{itemize}
			\item We show that, when no prior information about the reflector is available, the direction of position information is parallel to the reflector and independent of the \gls{Tx} and \gls{Rx} position. Hence, the line segment where the receiver can lie given the measurements of a \gls{NLOS} path, as identified in~\cite{MYJ07}, is always orthogonal to the reflecting surface.
			\item By encoding prior information about reflectors as prior distribution of the location of their corresponding \glspl{VA}, we study the effect of the accuracy of prior information on the intensity and direction of position information offered by single-bounce reflections.
		\end{itemize}

	\section{System Model}
		\label{sec:system model}
		
		\subsection{System Model}
\begin{figure}
			\centering
			\includegraphics[scale=1]{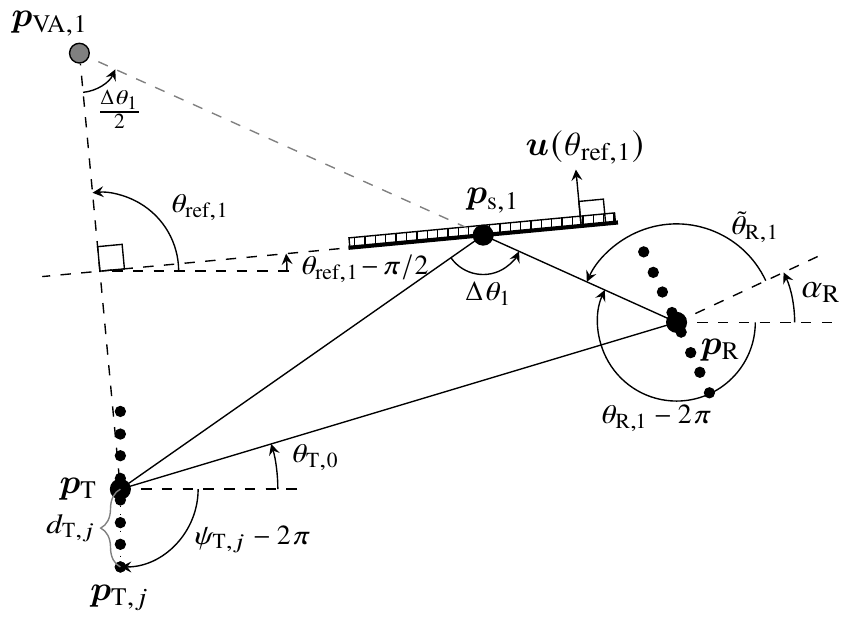}
			\sbox\glsscratchboxa{\footnotesize \glspl{ULA}}
			\sbox\glsscratchboxb{\footnotesize \gls{UCA}}
			\sbox\glsscratchboxc{\footnotesize \gls{Tx}}
			\sbox\glsscratchboxd{\footnotesize \gls{Rx}}
			\caption{Geometric model, example with \unhcopy\glsscratchboxa~at the \unhcopy\glsscratchboxc~and the \unhcopy\glsscratchboxd.}
			\label{fig:geometric_model}
		\end{figure}

		The \gls{Tx} consists of an array with $N_{\text{T}}$ antennas 
		and reference point located at
		$\vect{p}_{\text{T}} = [p_{\text{T,x}},\; p_{\text{T,y}}]\trans\in\mathbb{R}^2$, where $(\cdot)\trans$ denotes transposition. For the $j$-th element of the \gls{Tx} array, $d_{\text{T},j}$ and $\psi_{\text{T},j}$ are its distance and angle from the \gls{Tx} array's reference point as shown in Fig.~\ref{fig:geometric_model}.	The position of the $j$-th element of the \gls{Tx} array is given by
		$\vect{p}_{\text{T},j} = d_{\text{T},j}\vect{u}(\psi_{\text{T},j})\in\mathbb{R}^2, \quad j=0,\dots, N_{\text{T}} - 1$,
		where $\vect{u}(\psi) = [\cos(\psi), \; \sin(\psi)]\trans$.
		The \gls{Rx} consists of an array with $N_{\text{R}}$ antennas, 
		orientation $\alpha_{\text{R}}$ (with respect to the \gls{Tx} array's orientation)  and reference point located at  
		$\vect{p}_{\text{R}} = [p_{\text{R,x}},\; p_{\text{R,y}}]\trans\in\mathbb{R}^2$. The position of antenna elements at the \gls{Rx} array are defined similarly as for the \gls{Tx}. Between each \gls{Tx}-\gls{Rx} pair of antennas there are $L$ discrete propagation paths, where the first path ($l=0$) is the \gls{LOS} path 
		and the rest ($l=1,\ldots,L-1$) are single-bounce \gls{NLOS} paths. The $l$-th single-bounce \gls{NLOS} path results from a reflection on a flat surface with normal vector $\vect{u}(\theta_{\text{ref},l})$
		and point of incidence $\vect{p}_{\text{s},l} = [p_{\text{s},l,\text{x}},\; p_{\text{s},l,\text{y}}]\trans$. Each single-bounce \gls{NLOS} can be viewed as a direct path resulting from a \gls{VA}, with the $l$-th \gls{VA} located at $\vect{p}_{\text{VA},l} = [p_{\text{VA},l,\text{x}},\; p_{\text{VA},l,\text{y}}]\trans,\; l=1,\ldots,L-1$. The length of the $l$-th path is given by $d_l$, i.e. $d_l=\|\vect{p}_{\text{R}}-\vect{p}_{\text{T}}\|_2$ for $l=0$ and $d_l=\|\vect{p}_{\text{R}}-\vect{p}_{\text{VA},l}\|_2$ for $l\ne 0$, with $\|\cdot\|_2$ being the Euclidean norm.
		The \glspl{AOA} are defined as
		\begin{IEEEeqnarray}{rCl}
			\theta_{\text{R},l} &=& \begin{cases}
				\atan2(p_{\text{T},\text{y}} - p_{\text{R}, \text{y}}, p_{\text{T},\text{x}} - p_{\text{R}, \text{x}}), & l=0,\\
				\atan2(p_{\text{VA},l,\text{y}} - p_{\text{R},\text{y}}, p_{\text{VA},l,\text{x}} - p_{\text{R}, \text{x}}), & l\neq 0,		
			\end{cases}
			\IEEEeqnarraynumspace
		\end{IEEEeqnarray}
		with $\atan2\left(y,x\right)$ being the four-quadrant inverse tangent function. The \glspl{AOA} in the local frame of the \gls{Rx} are defined as $\tilde{\theta}_{\text{R},l} = \theta_{\text{R}, l} - \alpha_{\text{R}},\;l=0,\ldots,L-1$. With the observation that for a single-bounce reflection it holds that $\theta_{\text{ref}, l} = (\theta_{\text{T}, l} + \theta_{\text{R}, l})/2$, the \glspl{AOD} can be expressed as 
		\begin{IEEEeqnarray}{rCl}
			\theta_{\text{T},l} &=& \begin{cases}
				\theta_{\text{R},0} - \pi, &\hspace*{-0.05cm} l = 0,\\
				2 \atan2(p_{\text{VA},l,\text{y}}\hspace*{-0.03cm} - \hspace*{-0.03cm}p_{\text{T}, \text{y}}, p_{\text{VA},l,\text{x}} \hspace*{-0.03cm}-\hspace*{-0.03cm} p_{\text{R},l,\text{x}}) \hspace*{-0.03cm}-\hspace*{-0.03cm} \theta_{\text{R},l}, &\hspace*{-0.05cm} l\neq 0.
			\end{cases} 
			\IEEEeqnarraynumspace
		\end{IEEEeqnarray}
		
		The array dimensions are much smaller than the distances between \gls{Tx}, \gls{Rx} and  reflectors. Thus, the delay of the $l$-th path from \gls{Tx} element $j$ to \gls{Rx} element $i$ can be approximated by
		$ \tau_{l,i,j} \approx \tau_l - \tau_{\text{T},j}(\tilde{\theta}_{\text{T},l}) - \tau_{\text{R},i}(\tilde{\theta}_{\text{R},l})$, where $\tau_{\text{T},j}(\tilde{\theta}_{\text{T},l}) = d_{\text{T},j}\vect{u}\trans(\psi_{\text{T},j})\vect{u}(\tilde{\theta}_{\text{T},l})/c$, 
		$\tau_{\text{R},i}(\tilde{\theta}_{\text{R},l}) =  d_{\text{R},i}\vect{u}\trans(\psi_{\text{R},i})\vect{u}(\tilde{\theta}_{\text{R},l})/c$ and 
		\begin{IEEEeqnarray}{rCl}
			\tau_l &=& \begin{cases}
				(\|\vect{p}_{\text{R}}-\vect{p}_{\text{T}}\|_2 + d_{\text{clk}})/c, & l=0\\
				( \|\vect{p}_{\text{R}} - \vect{p}_{\text{VA},l}\|_2 + d_{\text{clk}})/c, & l\neq 0,
			\end{cases}\label{eq:pos2ch tau}
		\end{IEEEeqnarray} 
		where $d_{\text{clk}}=c \epsilon_{\text{clk}}$ with $\epsilon_{\text{clk}}$ as the clock offset between \gls{Tx} and \gls{Rx} and $c$ as the speed of light. 
		
		An \gls{OFDM} waveform with $N$ subcarriers and subcarrier spacing $\Delta f$ is considered. The set of used subcarriers is given by $\mathcal{P}$. A narrowband signal model is assumed, 
		i.e. $B/f_{\text{c}} \ll \lambda_{\text{c}}/D_{\max}$, where $B\approx \Delta f (\max(\mathcal{P}) - \min(\mathcal{P}))$ is the signal bandwidth, $f_{\text{c}}$ and $\lambda_{\text{c}}$ are the carrier frequency and wavelength, and $D_{\max}$ is the largest dimension of the \gls{Tx} and \gls{Rx} arrays. The received signal at the $p$-th subcarrier (for $p \in \mathcal{P}$) is 
		\begin{IEEEeqnarray}{rCl}
			\vect{y}[p] &=& \vect{m}[p] + \vect{\eta}[p],
			\label{eq:signal model}
			\IEEEeqnarraynumspace
		\end{IEEEeqnarray}
		where $\vect{\eta}[p] \sim \mathcal{N}_{\mathbb{C}}(\vect{0},\sigma_{\eta}^2\vect{I}_{N_{\text{R}}})$ is the \gls{AWGN} at the $p$-th subcarrier and
		\begin{IEEEeqnarray}{rCl}
			\vect{m}[p] &=& \sum_{l=0}^{L-1} h_l \e^{-\jj \omega_p \tau_l} \vect{a}_{\text{R}}(\tilde{\theta}_{\text{R},l}) \vect{a}_{\text{T}}\trans(\tilde{\theta}_{\text{T},l}) \vect{x}[p],\label{eq:m[p,b] definition}
			\IEEEeqnarraynumspace
		\end{IEEEeqnarray}
		where $h_l \in \mathbb{C}$ is the gain of the $l$-th path, $\omega_p = 2\pi p \Delta f$, and $\vect{x}[p]\in\mathbb{C}^{N_{\text{T}}}$  is the reference signal of the $p$-th subcarrier. With $\omega_{\text{c}} = 2\pi f_{\text{c}}$, the \gls{Tx} array steering vector $\vect{a}_{\text{T}}(\tilde{\theta}_{\text{T},l})$ is given
		\begin{IEEEeqnarray}{rCl}
			\vect{a}_{\text{T}}(\tilde{\theta}_{\text{T},l}) = [
			e^{\jj \omega_c \tau_{\text{T},1}(\tilde{\theta}_{\text{T},l})}, \;& \ldots, \;&\e^{\jj \omega_c \tau_{\text{T},N_{\text{T}}}(\tilde{\theta}_{\text{T},l})}
			]\trans\in\mathbb{C}^{N_{\text{T}}},
			\IEEEeqnarraynumspace
		\end{IEEEeqnarray}
		with the \gls{Rx} steering vector $\vect{a}_{\text{R}}(\tilde{\theta}_{\text{R},l}) $ defined similarly.

	\section{Cram\'er-Rao Lower Bound}
		\label{sec:Cramer-Rao lower bound}
		We first define the channel parameter vector $\vect{\phi}\in \mathbb{R}^{5L}$ as 
		\begin{IEEEeqnarray}{rCl}
			\vect{\phi} &=&  [\tau_0, \tilde{\theta}_{\text{T},0}, \tilde{\theta}_{\text{R},0},\vect{h}_0\trans,\cdots,\tau_{L-1}, \theta_{\text{T},L-1}, \theta_{\text{R},L-1},\vect{h}_{L-1}\trans,]\trans.
			\IEEEeqnarraynumspace
			\label{eq:position parameter vector definition}
		\end{IEEEeqnarray}
		and the position parameter vector 
		\begin{IEEEeqnarray}{rCl}
			\tilde{\vect{\phi}} &=& [\vect{p}_{\text{R}},\; \alpha_{\text{R}},\; d_{\text{clk}}, \vect{p}_{\text{VA},1},\cdots, \vect{p}_{\text{VA},L-1} ]\trans\in \mathbb{R}^{2L + 2}. 
			\IEEEeqnarraynumspace
		\end{IEEEeqnarray}
		According to the \gls{CRLB}, the covariance matrix $\vect{C}_{\hat{\tilde{\vect{\phi}}}}$  of any unbiased estimator $\hat{\tilde{\vect{\phi}}}$ of $\tilde{\vect{\phi}}$  satisfies  $\vect{C}_{\hat{\tilde{\vect{\phi}}}} - \vect{J}_{\tilde{\vect{\phi}}}^{-1} \succeq \vect{0}$~\cite{VT13}, where $\succeq \vect{0}$ denotes positive semi-definiteness and $\vect{J}_{\tilde{\vect{\phi}}}\in\mathbb{R}^{(2L+2)\times(2L+2)}$ is the hybrid \gls{FIM} of $\tilde{\vect{\phi}}$ given by $\vect{J}_{\tilde{\vect{\phi}}} = \vect{J}^{(\text{p})}_{\tilde{\vect{\phi}}} + \vect{J}^{(\text{o})}_{\tilde{\vect{\phi}}}\label{eq:J_phi definition},$
		with $\vect{J}^{(\text{p})}_{\tilde{\vect{\phi}}}$ and $\vect{J}^{(\text{o})}_{\tilde{\vect{\phi}}}$ accounting for the prior information and observation-related information on $\tilde{\vect{\phi}}$, respectively. We note that the hybrid FIM and the corresponding \gls{CRLB} characterize the estimation performance for a deterministic $\vect{\phi}$, where two sources of information are used: the received signal and the prior $p(\tilde{\vect{\phi}})$~\cite{WGK+18}. The observation-related FIM $\vect{J}^{(\text{o})}_{\tilde{\vect{\phi}}}$ can be obtained based on the \gls{FIM} $\vect{J}^{(\text{o})}_{\vect{\phi}}$ of the channel parameter vector $\vect{\phi}$ as $\vect{J}^{(\text{o})}_{\tilde{\vect{\phi}}} = \vect{T}\vect{J}^{(\text{o})}_{\vect{\phi}} \vect{T}\trans$. The entries of $\vect{J}^{(\text{o})}_{\vect{\phi}}\in \mathbb{R}^{5L \times 5L}$ and $\vect{T}\in \mathbb{R}^{2L+2 \times 5L}$ are given by
		\begin{IEEEeqnarray}{rCl}
			\hspace*{-0.5cm}\left[\vect{J}^{(\text{o})}_{\vect{\phi}}\right]_{i,j} &=& \frac{2}{\sigma_{\eta}^2}\sum_{p\in \mathcal{P}} \Re\bigg\{\frac{\partial \vect{m}\herm[p]}{\partial \phi_i} \frac{\partial \vect{m}[p]}{\partial \phi_j}\bigg\},\; i,j = 1,\ldots,5L,\\
			\hspace*{-0.5cm}\left[\vect{T}\right]_{i,j} &=& \partial\phi_j/\partial\tilde{\phi}_i, \; i = 1,\ldots, 2L + 2,\; j = 1,\ldots,5L,
			\label{eq:entries of channel parameter FIM LOS}
			\IEEEeqnarraynumspace
		\end{IEEEeqnarray}
		where $\vect{m}\herm$ is the conjugate transpose of $\vect{m}$ and $\Re\{\vect{m}\}$ is its real part. 
		Details on the required derivatives can be found in~\cite{KCS+19}.
		The \gls{PEB} for the \gls{Rx} is 
		defined as
		\begin{IEEEeqnarray}{rCl}
			\text{Rx PEB} &=& \sqrt{\big[\vect{J}_{\tilde{\vect{\phi}}}^{-1}\big]_{1,1} + \big[\vect{J}_{\tilde{\vect{\phi}}}^{-1}\big]_{2,2}}
			\IEEEeqnarraynumspace
		\end{IEEEeqnarray}
		and the \gls{PEB} for the \glspl{VA} is defined in a similar manner.

		The \gls{Rx} has prior information on the clock offset $p(\epsilon_{\text{clk}}') = \mathcal{N}(\epsilon_{\text{clk}}'; \epsilon_{\text{clk}}, \sigma_{\text{clk}}^2)$ and the \gls{VA}s' locations $p(\vect{p}_{\text{VA}, l}') = \mathcal{N}(\vect{p}_{\text{VA}, l}'; \vect{p}_{\text{VA}, l}, \Sigma_{\text{VA,pr},l})$, which encode map information about reflectors available at the \gls{Rx}, with $\mathcal{N}(\vect{x}; \vect{\mu}, \vect{\Sigma})$ denoting that $\vect{x}$ follows a Gaussian distribution with mean $\vect{\mu}$ and covariance $\vect{\Sigma}$. The hybrid FIM of the position parameter vector is
		\begin{IEEEeqnarray}{rCl}
			\vect{J}_{\tilde{\vect{\phi}}} &=& \vect{T}\vect{J}^{(\text{o})}_{\vect{\phi}}\vect{T}\trans +
			\begin{bmatrix}
				\vect{0} & \vect{0} \\
				\vect{0} & \vect{J}_{\text{VA,pr}}
			\end{bmatrix} +
			\frac{1}{c\sigma_{\text{clk}}^2}\vect{e}_4\vect{e}_4\trans,
			\label{eq:FIM position parameters}
		\end{IEEEeqnarray}
		where
		\begin{IEEEeqnarray}{rCl}
			\vect{J}_{\text{VA,pr}} &=& \begin{bmatrix}
				\vect{\Sigma}_{\text{VA,pr},1}^{-1} &  &  \vect{0} \\
				& \ddots & \\
				\vect{0} & & \vect{\Sigma}_{\text{VA,pr},L-1}^{-1}
			\end{bmatrix}
			\in \mathbb{R}^{2(L-1) \times 2(L-1)}
			\IEEEeqnarraynumspace
		\end{IEEEeqnarray}
		where $\vect{\Sigma}_{\text{VA,pr},l} \in \mathbb{R}^{2 \times 2}$ for $l=1,\cdots,L-1$ is the covariance matrix of the $l$-th \gls{VA}'s location given by  
		\begin{IEEEeqnarray}{rCl}
			\vect{\Sigma}_{\text{VA,pr},l} &\hspace*{-0.05cm}=\hspace*{-0.05cm}& [
			\vect{u}\left(\theta_{\text{R},l}\right) \;\hspace*{-0.05cm}
			\vect{u}_{\perp}\left(\theta_{\text{R},l}\right) 
			]
			\begin{bmatrix}
				\sigma_{l,\parallel}^{2} & \hspace*{-0.3cm}\rho_l \sigma_{l,\parallel} \sigma_{l,\perp} \\
				\rho_l \sigma_{l,\parallel} \sigma_{l,\perp} & \hspace*{-0.3cm} \sigma_{l,\perp}^{2}
			\end{bmatrix}
			\begin{bmatrix}
				\hspace*{-0.05cm}\vect{u}\trans\left(\theta_{\text{R},l}\right)\hspace*{-0.05cm} \\
				\hspace*{-0.05cm}\vect{u}\trans_{\perp}\left(\theta_{\text{R},l}\right) \hspace*{-0.05cm}
			\end{bmatrix}.\nonumber\\
			\IEEEeqnarraynumspace
		\end{IEEEeqnarray}
		where $\vect{u}_{\perp}(\theta) = \vect{u}(\theta - \pi/2)$.
		
		We employ the \gls{EFIM}~\cite{SW07}, to focus on the available information on the paramters of interest. Splitting $\vect{T}$ as $\vect{T} = 
		[\vect{T}_{\text{poc}}\trans\; \vect{T}_{\text{VA}}\trans
		]\trans$,
		with $\vect{T}_{\text{poc}} \in \mathbb{R}^{4 \times 5L}$ comprising the first four rows of $\vect{T}$ corresponding to the position and orientation parameters and clock offset and $\vect{T}_{\text{VA}}\in \mathbb{R}^{2(L-1) \times 5L}$ including the rest of the rows of  $\vect{T}$,
		the EFIM for the position and  orientation parameters and clock offset is given by 
		\begin{IEEEeqnarray}{rCl}
			\vect{J}_{\text{poc}} &\hspace*{-0.05cm}=\hspace*{-0.05cm}& \vect{T}_{\text{poc}} \vect{J}_{\vect{\phi}} \vect{T}_{\text{poc}}\trans \hspace*{-0.05cm}-\hspace*{-0.05cm} \vect{T}_{\text{poc}} \vect{J}_{\vect{\phi}} \vect{T}_{\text{VA}}\trans \vect{J}_{\text{VA}}^{-1}
			\vect{T}_{\text{VA}} \vect{J}_{\vect{\phi}} \vect{T}_{\text{poc}}\trans
			\hspace*{-0.05cm}+\hspace*{-0.05cm}\frac{\vect{e}_4 \vect{e}_{4}\trans}{(c\sigma_{\text{clk}})^2}, 
			\label{eq:J_poc}
			\IEEEeqnarraynumspace
		\end{IEEEeqnarray}
		where $\vect{J}_{\text{VA}} = 
		\vect{T}_{\text{VA}}
		\vect{J}_{\vect{\phi}}
		\vect{T}_{\text{VA}}\trans
		+
		\vect{J}_{\text{VA,pr}}$. 
		
		Making use of the fact that for large bandwidth and number of antennas the paths become asymptotically orthogonal~\cite{AZA+18}, $\vect{J}^{(\text{o})}_{\vect{\phi}}$ becomes a diagonal matrix. Indexing the diagonal elements of $\vect{J}^{(\text{o})}_{\vect{\phi}}$ by the parameter they correspond to, e.g. $J_{\tau_0} = [\vect{J}^{(\text{o})}_{\vect{\phi}}]_{1,1}$, it can be shown that \eqref{eq:J_poc} can be written as
		\begin{IEEEeqnarray}{rCl}
			\vect{J}_{\text{poc}} &=& \frac{J_{\tau_0}}{c^2} \vect{z}_{\tau_0} \vect{z}_{\tau_0}\trans +\frac{J_{\theta_{\text{T},0}}}{d_0^2} \vect{z}_{\theta_{\text{T},0}} \vect{z}_{\theta_{\text{T},0}}\trans  +
			\frac{J_{\theta_{\text{R},0}}}{d_0^2} \vect{z}_{\theta_{\text{R},0}} \vect{z}_{\theta_{\text{R},0}}\trans \nonumber \\
			&& + \sum_{l=1}^{L-1} \vect{J}_{l} +\frac{\vect{e}_4 \vect{e}_{4}\trans}{(c\sigma_{\text{clk}})^2} 
			\label{eq:J_poc_2}
			\nonumber
		\end{IEEEeqnarray}
		where the \gls{EFIM} $\vect{J}_l$ of the $l$-th \gls{NLOS} path is
		\begin{IEEEeqnarray}{rCl}
			\vect{J}_{l} =
			\frac{1}{\left|\vect{J}_{\text{VA},l}\right|}
			[
			\vect{z}_{\tau_l},\vect{z}_{\theta_{\text{T},l}},\vect{z}_{\theta_{\text{R},l}}  
			]
			\vect{M}_{l}
			[
			\vect{z}_{\tau_l},\vect{z}_{\theta_{\text{T},l}},\vect{z}_{\theta_{\text{R},l}}  
			]\trans
			\label{eq:J_l}
		\end{IEEEeqnarray}
		with 
		\begin{IEEEeqnarray}{rCl}
			\vect{z}_{\tau_l} &=& 	
			\begin{bmatrix} 
				-\vect{u}\trans\left(\theta_{\text{R},l}\right), & 0, & 1 	
			\end{bmatrix}\trans \label{eq:z_tau_l} \\
			\vect{z}_{\theta_{\text{T},l}} &=& 	
			\begin{bmatrix} 
				\vect{u}_{\perp}\trans\left(\theta_{\text{R},l}\right)
				, & 0, & 0	
			\end{bmatrix}\trans \label{eq:z_theta_Tl}\\
			\vect{z}_{\theta_{\text{R},l}} &=& 	
			\begin{bmatrix} 
				\vect{u}_{\perp}\trans\left(\theta_{\text{R},l}\right)
				, & -d_l, & 0	
			\end{bmatrix}\trans\label{eq:z_theta_Rl} \IEEEeqnarraynumspace
		\end{IEEEeqnarray}
		and $\vect{u}_{\perp}\left(\theta\right) =\vect{u}\left(\theta - \frac{\pi}{2}\right)$. The entries of $\vect{M}_{l} \in \mathbb{C}^{3 \times 3}$ and $|\vect{J}_{\text{VA}, l}|$ in \eqref{eq:J_l} are given in the Appendix.

	\section{Geometric Interpretation of Position Information}
		\label{sec:geometric interpretation of position information}
		It is interesting to carefully examine and obtain geometric intuition on the position information for the cases of perfect and no knowledge of the \gls{VA}'s position. The former case is straightforward: from \eqref{eq:J_l} for $\sigma_{l, \parallel}, \sigma_{l,\perp}\rightarrow 0$ we get
		\begin{IEEEeqnarray}{rCl}
			\vect{J}_l &=& \frac{J_{\tau_l}}{c^2} \vect{z}_{\tau_l} \vect{z}_{\tau_l}\trans +\frac{J_{\theta_{\text{T},l}}}{d_l^2} \vect{z}_{\theta_{\text{T},l}} \vect{z}_{\theta_{\text{T},l}}\trans  +
			\frac{J_{\theta_{\text{R},l}}}{d_l^2} \vect{z}_{\theta_{\text{R},l}} \vect{z}_{\theta_{\text{R},l}}\trans. \label{eq:NLOS FIM perfect prior}
		\end{IEEEeqnarray}
		As expected, in this case the \gls{NLOS} path acts in the same way as \gls{LOS} path. Using \eqref{eq:NLOS FIM perfect prior}  and \eqref{eq:z_tau_l}-\eqref{eq:z_theta_Rl} we can see that the rank of $\vect{J}_l$ is equal to 3, with each of the measurements providing position and orientation information independently: the \gls{TOA} provides position information in the radial direction, the \gls{AOD} and \gls{AOA} provide position information in the tangential direction and the \gls{AOA} provides orientation information.
		
\begin{figure}
			\centering
			\begin{adjustbox}{scale=0.85}
				\includegraphics[scale=1]{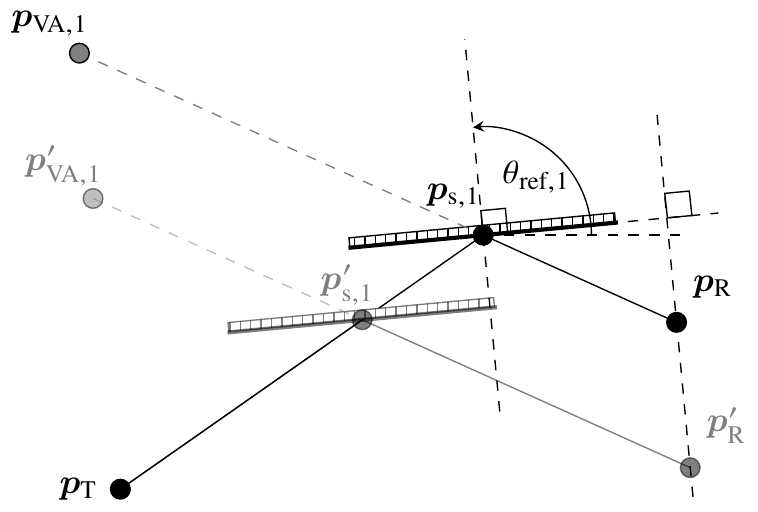}
			\end{adjustbox}
			\caption{Potential solutions for $p_{\text{R}}$, $p_{\text{s},l}$ and $p_{\text{VA}, l}$ explaining the measurements for a single-bounce reflection.}
			\label{fig:geometric interpretation}
		\end{figure}

		In the case of no knowledge of the \gls{VA}'s location, i.e. $\sigma_{l, \parallel}, \sigma_{l,\perp} \rightarrow \infty$, it can be shown that \eqref{eq:J_l} becomes
		\begin{IEEEeqnarray}{rCl}
			\vect{J}_{l} &=& j_l \vect{z}_l \vect{z}_l\trans\label{eq:NLOS FIM no prior}	\IEEEeqnarraynumspace
		\end{IEEEeqnarray}
		where 
		\begin{IEEEeqnarray}{rCl}
			j_l &=&  \frac{J_{\tau_l} J_{\theta_{\text{T},l}} J_{\theta_{\text{R},l}}}{| \vect{J}_{\text{VA},l}| c^2 d_l^2 d_{\text{T,s},l}^2 \; \cos^2(\Delta \theta_{l}/2)} \\
			\vect{z}_{l} 
			&=&[
			\vect{u}_{\perp}\trans(\theta_{\text{ref},l}), \; -d_{\text{R,s},l}\; \cos(\Delta \theta_{l}/2), \; \sin(\Delta \theta_{l}/2) 	
			]\trans, \label{eq:z_l no prior}
			\IEEEeqnarraynumspace
		\end{IEEEeqnarray}
		with $d_{\text{T,s}, l} = \|\vect{p}_{\text{s},l} - \vect{p}_{\text{T}}\|_2$ and $d_{\text{R,s}, l} = \| \vect{p}_{\text{R}} - \vect{p}_{\text{s},l} \|_2$.
		We can observe from~\eqref{eq:NLOS FIM no prior} that, as first noted in~\cite{MWB+19}, $\vect{J}_l$ has rank $1$. Furthermore, from~\eqref{eq:z_l no prior} and Fig.~\ref{fig:geometric_model}, we conclude that the direction of position information is always parallel to the reflecting surface and independent of the \gls{Tx} and \gls{Rx} location. At first glance, this is a surprising result, since for \gls{LOS} paths and \gls{NLOS} paths with perfect knowledge of the corresponding \glspl{VA}' location the direction of position information depends on $\vect{p}_{\text{R}}$ and $\vect{p}_{\text{T}}$. A geometrically intuitive explanation of this result can be obtained from Fig.~\ref{fig:geometric interpretation}. 
		In Fig.~\ref{fig:geometric interpretation} we consider a single-bounce reflection and plot two potential geometries $\{\vect{p}_{\text{R}},\vect{p}_{s, 1}, \vect{p}_{\text{VA},1}\}$ and $\{\vect{p}_{\text{R}}',\vect{p}_{s, 1}', \vect{p}_{\text{VA},1}'\}$ that would produce the same \gls{TOA}, \gls{AOD} and \gls{AOA}. In fact, there are infinitely many such geometries, parametrized as
		\begin{IEEEeqnarray}{rCl}
			\vect{p}_{\text{R}} 
			&=& \vect{p}_{\text{T}} - c\cdot \tau_l \vect{u}(\theta_{\text{R}, l}) +  2 \lambda \cos(\Delta\theta_l/2)\vect{u}(\theta_{\text{ref}, l}) \label{eq:p_R locus} \\
			\vect{p}_{\text{VA}, l} &=& \vect{p}_{\text{T}} +  2 \lambda \cos(\Delta\theta_l/2)\vect{u}(\theta_{\text{ref}, l}) \\
			\vect{p}_{\text{s}, l} &=& \vect{p}_{\text{T}} + \lambda \vect{u}(\theta_{\text{T}, l}), \quad 
		\end{IEEEeqnarray}
		with $0<\lambda< c \cdot \tau_l$. As can be seen in \eqref{eq:p_R locus}, the locus of $\vect{p}_{\text{R}}$ is a line segment normal to the reflecting surface. Hence, the \gls{NLOS} path associated with the reflection provides position information only in the direction that is perpendicular to this line segment, i.e. in the direction parallel to the reflecting surface. An implication of this result is that information from single-bounce paths from parallel (or close to parallel) reflecting surfaces may not suffice for target localization.
			
	\section{Numerical Results}
		\label{sec:numerical results}
		\subsection{Simulation setup}
		We set $f_{\text{c}} = \SI{38}{\giga\hertz}$, $N = 1024$, $\Delta f = \SI{120}{\kilo\hertz}$, $\mathcal{P} = \{-420,\ldots,-1,1\ldots, 420\}$ and $B \approx \SI{100}{\mega\hertz}$. 
		The entries of $\vect{x}[p]$ have constant amplitude and random phase, with $\mathrm{E}[\|\vect{x}[p]\|_2^2] = \SI{0}{\dBm}$. The noise variance is $\sigma_{\eta}^2 =  10^{0.1 (n_{\text{Rx}} + N_0)} N \Delta f $, where $N_0=\SI{-174}{\dBm\per\hertz}$ is the noise power spectral density and $n_{\text{Rx}} = \SI{8}{\decibel}$ is the \gls{Rx} noise figure. 
		
\begin{figure}
			\centering
			\begin{adjustbox}{scale=0.90}
				\includegraphics[scale=1]{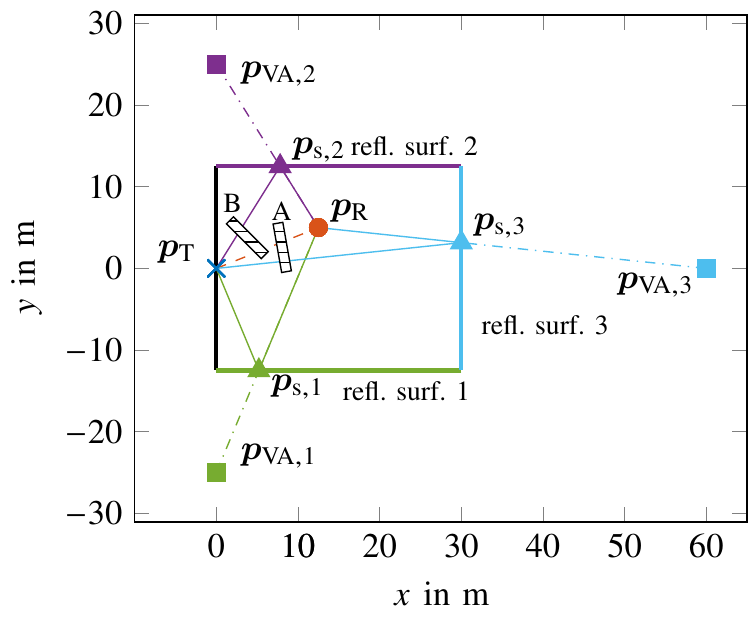}
			\end{adjustbox}
			\caption{Simulation scenario.}
			\label{fig:simulation_scenario}
		\end{figure}

		We consider the scenario depicted in Fig.~\ref{fig:simulation_scenario}, where the \gls{Tx} lies at the origin and the \gls{Rx} at $\vect{p}_{\text{R}} = [12.5, 5]\trans\si{\meter}$. The \gls{Tx} has a \gls{ULA} with 32 antennas and the \gls{Rx} has a \gls{UCA} with 16 antennas and orientation $\alpha_{\text{R}}$. The \glspl{VA} resulting from single-bounce reflections at the rooms' walls are located at $p_{\text{VA}, 1} = [0,-25]\trans \si{\meter}$, $p_{\text{VA}, 2} = [0,25]\trans \si{\meter}$ and $p_{\text{VA}, 3} = [60,0]\trans \si{\meter}$.
		In order to concentrate on the potential implications of the results presented in Sec.~\ref{sec:geometric interpretation of position information}, we assume that the \gls{Rx} orientation $\alpha_{\text{R}}$ is known and the \gls{Tx} and \gls{Rx} are perfectly synchronized. We consider two \gls{NLOS}-only cases:
		\begin{itemize}
			\item \emph{case A}: the paths corresponding to the \nth{1} and \nth{2} \glspl{VA} are received;
			\item \emph{case B}: the paths corresponding to the \nth{1} and \nth{3} \glspl{VA} are received.
		\end{itemize} 
		The amplitude of the complex path gain of the $l$-th path is $|h_l| = \sqrt{\gamma_l} \lambda/(4\pi d_l)$, where $\gamma_l=0.1\;\forall l$, is the reflection coefficient, and the phase is uniformly distributed. 
		
		\subsection{Results}
\begin{figure}
			\centering
			\subfloat[Eigenvalues of $\vect{J}_1$.]{\begin{adjustbox}{scale=0.97}
					\includegraphics[scale=1]{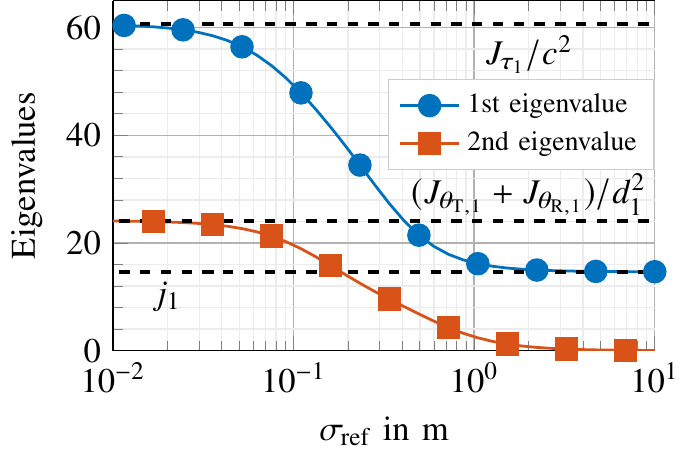}
			\end{adjustbox}}%
			
			\subfloat[Direction eigenvectors of $\vect{J}_1$.]{\begin{adjustbox}{scale=0.97}
					\includegraphics[scale=1]{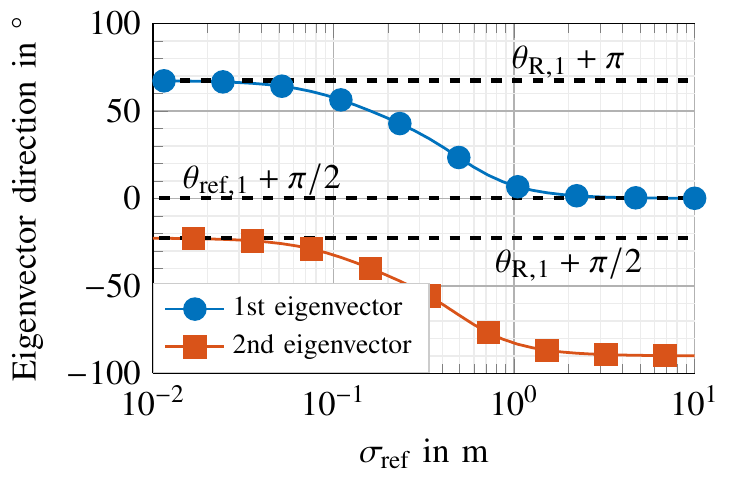}
			\end{adjustbox}}%
			\caption{Eigenvalues and directions of eigenvectors of the \gls{EFIM} $\vect{J}_1$  of \gls{VA} 1 as function of the prior \gls{VA} position error $\sigma_{\text{ref}}$.}
			\label{fig:eigenvalues_and_directions_vs_sigma_refl}
		\end{figure}

		From the analysis in Sec.~\ref{sec:geometric interpretation of position information} we have a clear picture about the position information offered by single-bounce \gls{NLOS} paths under perfect or no prior knowledge of their corresponding \glspl{VA} locations. 
		In order to gain more insight about the intermediate cases, setting $\rho_1=0$ and $\sigma_{1, \parallel} = \sigma_{1,\perp} = \sigma_{\text{ref}}/\sqrt{2}$, we plot the eigenvalues and the directions of the eigenvectors for varying $\sigma_{\text{ref}}$ in Fig.~\ref{fig:eigenvalues_and_directions_vs_sigma_refl}.
		We see that, as expected, when knowledge about the \gls{VA}'s position is accurate ($\sigma_{\text{ref}} \rightarrow 0$), for known orientation and perfect synchronization (in this case $\vect{z}_{\theta_{\text{T}, l}} = \vect{z}_{\theta_{\text{R}, l}}=\vect{u}_{\perp}\trans(\theta_{\text{R}, l})$), $\vect{J}_1$ has two strong eigenvalues, with the eigenvectors pointing in the radial  and the tangential direction. As $\sigma_{\text{ref}}$ increases, the strongest eigenvalue decreases, starting from 
		$J_{\tau_1}/c^2$
		and converges to $j_1$, while the second eigenvalue vanishes, resulting in a rank-1 $\vect{J}_1$.
		The direction of the eigenvector corresponding to the strongest eigenvalue gradually changes from $\theta_{\text{R}, 1} +\pi$, which corresponds to range information, to $\theta_{\text{ref}, 1} + \pi/2$, that is parallel to the reflecting surface.
		
\begin{figure}
			\centering
			\begin{adjustbox}{scale=0.97}
				\includegraphics[scale=1]{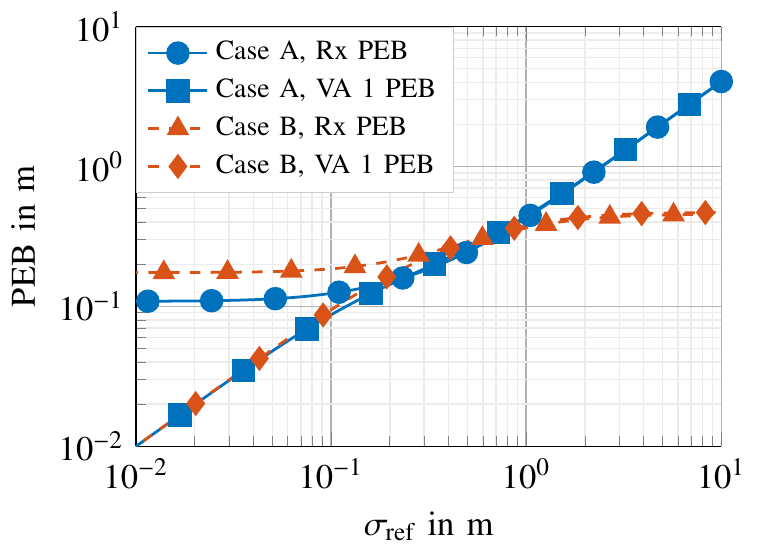}
			\end{adjustbox}
			\caption{\gls{Rx} and \gls{VA} 1 \gls{PEB} as function of the prior \gls{VA} position error $\sigma_{\text{ref}}$.}
			\label{fig:Rx_and_VA_PEB_vs_sigma_refl}
		\end{figure}

		In Fig.~\ref{fig:Rx_and_VA_PEB_vs_sigma_refl} we plot the \gls{PEB} of the \gls{Rx} and \gls{VA} 1 for the two considered cases as functions of $\sigma_{\text{ref}}$. We set again $\rho_l=0$ and $\sigma_{l, \parallel} = \sigma_{l,\perp} = \sigma_{\text{ref}}/\sqrt{2}, \; l=1,2,3$.
		We see that for $\sigma_{\text{ref}} \rightarrow 0$ the \gls{PEB} of \gls{VA} 1 converges to 0, while the \gls{Rx} \gls{PEB} converges to its lowest value as the two paths behave as \gls{LOS} paths, providing position information in linearly independent directions. In case A, 
		as $\sigma_{\text{ref}}$ increases the two paths provide position information in almost the same direction, as they arise from parallel reflecting surfaces, with the available information in the orthogonal direction decreasing with increasing $\sigma_{\text{ref}}$. 
		As a result, 
		for high values of $\sigma_{\text{ref}}$ (i.e. less accurate prior) the \gls{PEB} of the \gls{Rx} and \gls{VA} 1 grows linearly with $\sigma_{\text{ref}}$. 
		In case A, for high values of $\sigma_{\text{ref}}$ (i.e. less accurate prior) the \gls{PEB} of the \gls{Rx} and \gls{VA} 1 grows linearly with $\sigma_{\text{ref}}$. This is due to the fact that the two paths provide position information in almost the same direction, as they arise from parallel reflecting surfaces, with the available information in the orthogonal direction decreasing with increasing $\sigma_{\text{ref}}$. For moderate values of $\sigma_{\text{ref}}$ ($<\SI{1}{\meter}$), good positioning accuracy is achievable as, the directions of the strongest eigenvectors of $\vect{J}_1$ and $\vect{J}_2$ are sufficiently distinct.
		On the contrary, in case B, the \gls{PEB} of the \gls{Rx} and \gls{VA} 1 saturates for high values of $\sigma_{\text{ref}}$, as the two paths provide position information in different directions, resulting from the two perpendicular walls. 
		Therefore, combining the two \gls{NLOS} paths the \gls{Rx} position can be resolved and, 
		consequently, the position of the \gls{VA}.

	\section{Conclusion}
		\label{sec:conclusion}
		We provided an analysis of the Fisher information on position, orientation and clock offset provided by single-bounce \gls{NLOS} paths. The effect of prior map information on the position information was studied. It was shown that when no prior knowledge is available, the direction of position information is always parallel to the reflecting surface, independent of the \gls{Rx} target position. We also provided a geometrically intuitive explanation of the result. Numerical examples considering a practical room geometry showed that, as a consequence of the aforementioned analysis, the availability of different multipath components can have a significant impact on the achievable positioning accuracy.

	\appendix[Entries of \texorpdfstring{$\vect{M}_l$}{M\_l} and \texorpdfstring{$|\vect{J}_{\text{VA}, l}|$}{J\_{VA,l}} in \eqref{eq:J_l}]
		The entries of $\vect{M}_l$ and $|\vect{J}_{\text{VA}, l}|$ in \eqref{eq:J_l} are given by
		\begin{IEEEeqnarray}{rCl}
		\left[\vect{M}_{l}\right]_{1,1} &\hspace*{-0.02cm}=\hspace*{-0.02cm}& \frac{J_{\tau_l}}{c^2} J_{\theta_{\text{T},l}} A^2 F \hspace*{-0.02cm}+\hspace*{-0.02cm} \frac{1}{(1-\rho_l^2)\sigma_{l, \parallel}^2}\Bigg(J_{\theta_{\text{T},l}} B^2 \hspace*{-0.02cm}+\hspace*{-0.02cm} \frac{J_{\theta_{\text{R},l}}}{d_{l}^2} \hspace*{-0.02cm}+\hspace*{-0.02cm} \frac{G}{\sigma_{l,\perp}^{2}}\Bigg) \nonumber \\
		\left[\vect{M}_{l}\right]_{2,2} &\hspace*{-0.02cm}=\hspace*{-0.02cm}& \frac{J_{\theta_{\text{T},l}}}{d_{l}^2} \Bigg( \frac{J_{\tau_l}}{c^2} F \hspace*{-0.02cm}+\hspace*{-0.02cm} \frac{1}{(1-\rho_l^2)\sigma_{l,\parallel}^2} P \Bigg) \nonumber \\
		\left[\vect{M}_{l}\right]_{3,3} &\hspace*{-0.02cm}=\hspace*{-0.02cm}& \frac{J_{\theta_{\text{R},l}}}{d_{l}^2}\Bigg(\hspace*{-0.02cm}
		J_{\theta_{\text{T},l}} B^2 Q \hspace*{-0.02cm}\hspace*{-0.02cm}+\hspace*{-0.02cm} \frac{1}{(1\hspace*{-0.02cm}-\hspace*{-0.02cm}\rho_l^2)\sigma_{l,\perp}^2} \Bigg( \hspace*{-0.02cm}	\frac{J_{\tau_l}}{c^2} \hspace*{-0.02cm}+ \hspace*{-0.02cm}J_{\theta_{\text{T},l}} A \hspace*{-0.02cm}+ \hspace*{-0.02cm}\frac{G}{\sigma_{l,\parallel}^{2}} \hspace*{-0.02cm} \Bigg)	\Bigg) \nonumber \\
		\left[\vect{M}_{l}\right]_{1,2} &\hspace*{-0.02cm}=\hspace*{-0.02cm}& \left[\vect{M}_{l}\right]_{2,1} \hspace*{-0.02cm}=\hspace*{-0.02cm} \frac{J_{\tau_l}}{c^2} \frac{J_{\theta_{\text{T},l}}}{d_{l}} \Bigg(A F \hspace*{-0.02cm}+\hspace*{-0.02cm} B \frac{\rho_l}{(1\hspace*{-0.02cm}-\hspace*{-0.02cm}\rho_l^2)\sigma_{l,\parallel}^2 \sigma_{l,\perp}^2} \Bigg)
		\nonumber \\
		\left[\vect{M}_{l}\right]_{1,3} &\hspace*{-0.02cm}=\hspace*{-0.02cm}& \left[\vect{M}_{l}\right]_{3,1} \hspace*{-0.02cm}=\hspace*{-0.02cm}  \frac{J_{\tau_l}}{c^2}  \frac{J_{\theta_{\text{T},l}}}{d_{l}^2} \Bigg(\frac{\rho_l}{(1\hspace*{-0.02cm}-\hspace*{-0.02cm}\rho_l^2)\sigma_{l,\parallel}^2 \sigma_{l,\perp}^2} \hspace*{-0.02cm}-\hspace*{-0.02cm} J_{\theta_{\text{T},l}} A B \Bigg) \nonumber \\
		\left[\vect{M}_{l}\right]_{2,3} &\hspace*{-0.02cm}=\hspace*{-0.02cm}& \left[\vect{M}_{l}\right]_{3,2} \hspace*{-0.02cm}=\hspace*{-0.02cm} \frac{J_{\theta_{\text{T},l}}}{d_l} \frac{J_{\theta_{\text{R},l}}}{d_{l}^2}
		\Bigg(B Q \hspace*{-0.02cm}+\hspace*{-0.02cm} \frac{ A \rho_l}{(1 \hspace*{-0.02cm}-\hspace*{-0.02cm} \rho_l^2)\sigma_{l,\parallel}^2 \sigma_{l,\perp}^2} \Bigg)\nonumber\\
		|\vect{J}_{\text{VA}, l}| &\hspace*{-0.02cm}=\hspace*{-0.02cm}& \frac{J_{\tau_l}}{c^2} \big( J_{\theta_{\text{T},l}} B^2 \hspace*{-0.02cm}+\hspace*{-0.02cm} F\big) \hspace*{-0.02cm} +\hspace*{-0.02cm} \frac{P + J_{\theta_{\text{T},l}}B (B + 2\rho_l A \sigma_{l,\parallel}/\sigma_{l,\perp})}{(1 \hspace*{-0.02cm}-\hspace*{-0.02cm} \rho_l^2)\sigma_{l,\parallel}^2} \nonumber\\
		&&\hspace*{-0.02cm}+\hspace*{-0.02cm} J_{\theta_{\text{T},l}} A^2 F, 
		\IEEEeqnarraynumspace\nonumber
		\end{IEEEeqnarray}
		where $A=\tan\big(\frac{\Delta \theta_{l}}{2} \big)/d_{\text{T,s},l}$,
		$B=1/d_{l} - 1/d_{\text{T,s},l}$, $\Delta \theta_{l}=\theta_{\text{R},l}-\theta_{\text{T},l}$, $F=\frac{J_{\theta_{\text{R},l}}}{d_l^2}+\frac{1}{(1-\rho_l^2)\sigma_{l,\perp}^{2}}$, $G= 1 +2 \rho_l \sigma_{l,\parallel} \sigma_{l,\perp} J_{\theta_{\text{T},l}} A B$, $P=\frac{J_{\theta_{\text{R},l}}}{d_l^2}+\frac{1}{\sigma_{l,\perp}^{2}}$ and $Q=\frac{J_{\tau_{l}}}{c^2}+\frac{1}{(1-\rho_l^2)\sigma_{l,\parallel}^{2}}$.
		\bibliographystyle{IEEEtran}
		\bibliography{IEEEabrv,References}
	
\end{document}